\documentclass{aastex}
\usepackage{emulateapj5,apjfonts,psfig}

\def\xmm{{\sl XMM-Newton }}
\def\asca{{\sl ASCA }}

\def\chandra{{\sl Chandra }}
\def\ergsec{\hbox{erg s$^{-1}$ }}
\def\ergcm{\hbox{erg cm$^{-2}$ s$^{-1}$ }}

\def\Msun{$M_{\odot}$ }
\def\Rsun{$R_{\odot}$ }

\def\it{\sl}

\slugcomment{Accepted for publication in The Astrophysical Journal Letters}

\shorttitle{X-RAY EMISSION LINES FROM VELA X-1}

\shortauthors{SCHULZ et al.}

\begin{document}

\title{The Ionized Stellar Wind in Vela X-1 During Eclipse}
\author{
Norbert. S. Schulz\altaffilmark{1},
Claude. R. Canizares\altaffilmark{1}
Julia. C. Lee\altaffilmark{1}
and
Masao Sako\altaffilmark{2}
 }
\altaffiltext{1}{Center for Space Research, Massachusetts Institute of Technology,
Cambridge, MA 02139, nss, crc, jlee@space.mit.edu.}
\altaffiltext{2}{Columbia Astrophysics Laboratory and Department of Physics, Columbia
University, New York 10027, masao@astro.columbia.edu.}

\begin{abstract}
We present a first analysis of a high resolution X-ray spectrum of the
ionized stellar wind of Vela X-1 during eclipse. The data were obtained with the
High Energy Transmission Grating Spectrometer onboard the \chandra X-ray Observatory.
The spectrum is resolved into emission lines with fluxes between
0.02 and 1.04$\times 10^{-4}$ photons cm$^{-2}$s$^{-1}$. We identify lines from a variety of
charge states, including fluorescence lines from cold material, a warm photoionized
wind. We can exclude signatures from collisionally ionized plasmas. For the first time
we identify fluorescence lines from L-shell ions from lower Z elements.  
We also detect radiative recombination continua from a kT = 10 eV (1.2$\times 10^5$ K)
photoionized optically thin gas. The fluorescence line fluxes infer the existence
of optically thick and clumped matter within or outside the warm photoionized plasma.
\end{abstract}


\section{Introduction}

Massive early type stars show substantial
mass loss through a strong wind. Matter is radially accelerated via
momentum transfer from the stellar UV radiation field through absorption and scattering
in resonance lines of abundant ions (Castor, Abott, and Klein 1975).
In general the stellar wind
itself is a source of X-rays, which form through reverse shocks generated by high density
material in the wind (Lucy and White 1980, Owocki et al. 1988, Feldmeier et al. 1997).
Typical observed X-ray luminosities range from a few times 10$^{31}$ \ergsec for
early type B-stars to a few times 10$^{32}$ \ergsec for massive O-stars (Chlebowski
et. al 1989, Bergh\"ofer et al. 1995). Recent \chandra and \xmm
observations resolved the X-ray emission of several early type stars confirming
the existence of collisionally ionized plasmas, but also suggested the possibility that
alternative emission mechanisms for the observed X-ray emission need to be investigated.
(Schulz et al. 2000, Kahn et al. 2001, Waldron and Cassinelli 2001).

The presence of a neutron star companion leads to a typical Bondi-Hoyle accretion
flow (Bondi and Hoyle 1944), where a fraction
of the stellar wind gets swept up by the neutron star, which generates a strong X-ray continuum.  
The X-ray luminosity of such a source is many orders of magnitudes larger than
the one expected from shocks in the stellar wind, and X-rays from that source
are reprocessed in the wind resulting in discrete emission lines (Stevens
and Kallman 1990).

A prototype of such a system is the high-mass X-ray binary (HMXB) Vela X-1 (4U 0900-40).
Its early type primary star HD 77581 is a massive B 0.5 Ib type supergiant (Brucato and
Kristian 1972) with a mass and radius of 34 \Rsun and 23 \Msun (from Nagase 1989), and an inferred
mass loss rate of $> 10^{-7}$ \Msun yr$^{-1}$ (Sato et al. 1986, Sako et al. 1999). The terminal
wind velocity has been determined to be 1700 km s$^{-1}$ (Dupree et al. 1980). The neutron star
exhibits 282 s X-ray pulsations (Forman et al. 1973) and
orbits the center of mass of the system at a distance of only about 0.6 stellar radii from the
surface of the supergiant (a = 53.4 \Rsun) with a period of 8.96 days. This means that
the neutron star is deeply embedded within the
acceleration zone of the stellar wind.

The X-ray continuum of Vela X-1 has been studied extensively in X-rays. The general shape
can be explained by a power law with a high energy cutoff typical for X-ray pulsars (White,
Swank, and Holt 1983). Its average X-ray luminosity is 
$\sim 10^{36}$ \ergsec, which is consistent with accretion from a stellar wind without
Roche lobe overflow (Conti 1978). 

In this paper we are primarily concerned with X-ray
line emission from Vela X-1 during the middle of its eclipse.
A first analysis of a moderate resolution X-ray line spectrum of Vela X-1 during
eclipse was performed on a set observations
with \asca, which covered an entire eclipse transition from ingress to egress (Nagase et al. 1994).
The observations revealed lines from He- and H-like ions species.
It was suggested that the He-like
transitions are formed by cascades following radiative recombination. Sako et al. 1999
identified recombination lines and
radiative recombination continua produced by photoionization in an extended stellar wind.
The deduced differential emission measure distribution was consistent
with the wind model by Hatchett $\&$ McCray (1977).
Sako et al. also indictated the existence of fluorescence lines from ion species of lower Z than Fe. 

\section{Chandra Observations and Data Reduction}

Vela X-1 was observed with the HETGS on 2000 April 13th (09:57:52 UT) continuously for 28 ks.
The HETGS carries 2 different types of transmission
gratings, the Medium Energy Gratings (MEG) and the High Energy Gratings (HEG).
It allows for high resolution spectroscopy between
about 1 and 35 \AA ~with a peak spectral resolution at 12 \AA ~of $\lambda/\Delta\lambda \sim 1400$
and at 1.8 \AA ~of $\lambda/\Delta\lambda \sim 180$ in 1st order HEG.
The dispersed spectra were recorded with an array of 6 charged coupled devices (CCDs).
We refer to the available Chandra X-ray Center (CXC) docments for more detailed
descriptions of the spectroscopic instruments\footnote{http://asc.harvard.edu/udocs/docs/docs.html}.

We recorded a total of 3200 events in the co-added 1st order MEG and 1700 events in the HEG
after standard grade selection. The CXC provided aspect corrected event lists via
standard pipeline processing.
These data were reprocessed using the latest available data processing products.
The wavelength scale has a zero point accuracy of 0.002 \AA ~in MEG and 0.001 \AA ~in HEG
1st order. The current status of the overall wavelength calibration is better than 0.05$\%$.
For the line analysis of the high resolution spectra we divide first by the instrument effective area and exposure, 
which is legitimate, because the spectral bin size is larger than the actual response of the instrument.

The image in the 0th order provides a source position as well as a medium resolution X-ray spectrum.
The position of HD 77581 is well known to a precision of about 0.3 arcsec (Roeser et al. 1991).
After correcting for a 2" aspect drift we find the position of the X-ray source to be within
within 0.5" of this position. We also find no other X-ray sources within a radius of 30" around Vela X-1
with a flux above 1.2$\times 10^{-15}$\ergcm.

The CCD spectrum of that image also allows us to draw comparisons with previous \asca observations,
which are of similar spectral resolution, but cover a larger fraction of the eclipse.
This CCD spectrum is only moderately piled up (~7$\%$) and can be used to determine
qualitative characteristics of the observed emission as well as the total observed luminosity.
The continuum model is the same as used by Sako et al. (1999), which consists of two absorbed power laws from 
the scattered component and a direct component with indices fixed to 1.7. The choice of this
model is only motivated by the fact that we can draw comparisons with specifically 
the \asca results. Since the spectrum shows significant line emission and we added several gaussian functions  
to fit the brighter lines observed in the \asca spectrum (see table 2 in Sako et al. 1999).

In general we find good agreement with the \asca results. The most striking  difference
is the column density of the direct component, which is twice as high and also indicates a much fainter
direct component. This is a consequence of the fact that the \chandra observations exposure covers a
smaller range in orbital phase around phase zero. We verfied this by re-analysing the \asca data using the \chandra phase interval.
From an X-ray flux of 0.88$\times 10^{-11}$\ergcm  we then deduce 2.1$\times 10^{33}$ \ergsec for the luminosity (0.5-10 keV)
assuming a distance of 1.9 kpc (Sadakane et al. 1985).

\section{The HETGS Spectra}

The spectral continuum is well modeled in the zero order spectrum and we use it to subtract
the continuum from the summed and background subtracted first order grating data.
Figure 1 shows the residual spectrum between 3.1 and 12.5 ~\AA~. 

\subsection{Fluorescence Emission}

The spectrum in Figure 1 shows a variety of fluorescent lines (green model) 
from Mg, Si, S, Ar, Ca, and Fe. The line properties, including possible
identifications, are listed in Table 1. For identifications we use the adjusted 
calculations from House (1969). A few are also tabulated in Mewe (1994).
The measured lines do not show significant velocity shifts or broadening and can be generally
identified with near neutral charge states. For the elements 
Fe, Ca, and Ar we cannot distinguish between different charge states and here
table 1 lists the most likely range based on the wavelengths given in House (1969).
In the case of Mg, Si, and S we can identify various states, however for
the lowest charge state we can again only identify a range of most likely ionization states.  

\begin{figure*}[t]
\centerline{\epsfxsize=16.5cm\epsfbox{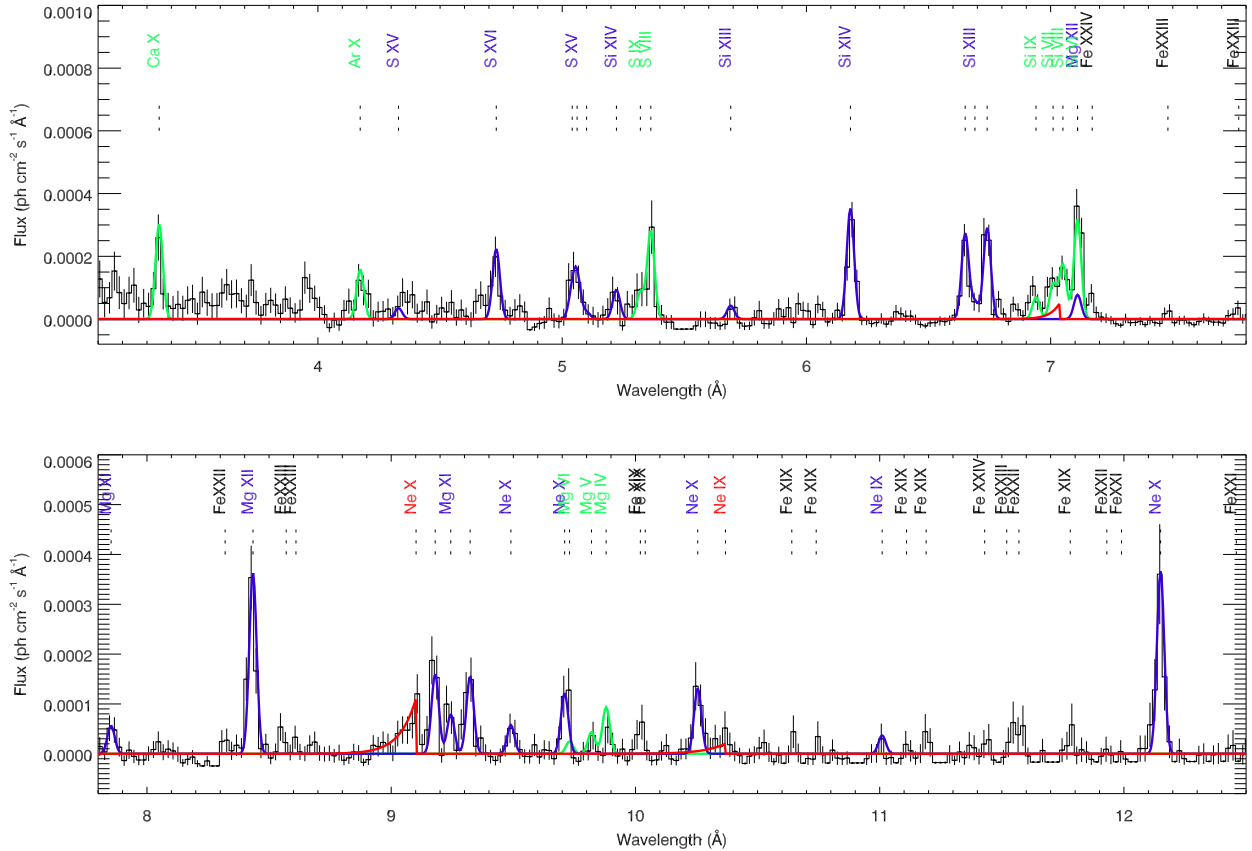}}
\figcaption{The residual HETGS spectrum in 0.02 ~\AA~ bin with identified lines. Fluorescence lines
are plotted in green. 
Highly ionized lines (blue) and RRC features (red) indicate an optically thin photoionized plasma.
\label{counts}}
\end{figure*}

For each detected line, we calculate an equivalent hydrogen column density
required to produce the observed line fluxes using the method described in
Sako et al.\ (1999).  Fluorescent yields and photoionization cross sections
are adopted from Kaastra and Mewe (1993) and Verner et al.\ (1996),
respectively. Kaastra and Mewe (1993) predict an Fe K$\beta$/K$\alpha$ flux ratio
of $\sim 0.13$. We fix the width of the K$\beta$ line to that of the K$\alpha$
line and infer a ratio of $0.124 \pm 0.095$. Since this consistent with the
theoretical value we assume that the iron line flux, and, hence the estimated 
column density, is not affected by resonant Auger destruction (Ross, Fabian,
and Brandt 1996). Resonant Auger destruction occurs in regions, where line optical
depths are significant. A fluorescent photon from the K-shell can be 
resonantely absorbed by ions that have vacancies in the L-shell. However, due
to high autoionization yields in atoms of much lower Z than Fe, the L-shell ion will
most likely Auger decay. In table 1 the quantity $Y_{\rm eff}$ denotes the
fraction of the Fe K$\alpha$ column density for each identified charge state
of the other elements.  
For Fe K $\alpha$ we derive a column density of 3.3$\times 10^{22}$ cm$^{-2}$.
If the $Y_{\rm eff}$s are added up for each element we observe a correlation of 
the column density with $Z$, which could indicate the process of resonant Auger 
destruction at work. Such an interpretation was also given by Sako et al.\ (1999).
However, for Si such an interpretation is problematic, 
since the Si VII and Si VIII lines appear exceptionally strong braking that correlation. 
In this respect the detection of fluorescent lines from L-shell ions suggests 
a somewhat modified explanation for the observed behavior.

\vbox{
\footnotesize
\begin{center}
{\sc TABLE 1\\
FLOURESCENT LINE ENERGIES AND FLUXES.}
\vskip 4pt
\begin{tabular}{lcccccc}
\hline
\hline
     Ion & type$^1$ & $\lambda_{exp}^1$ & $\lambda_{meas}$ & Flux$^2$ & Y$_{eff}^3$\\
         &  &\AA & \AA &  & & \\
\hline
     & & & & \\
     Fe & K $\beta$  & 1.756 & $1.748\pm0.020$ & $1.29\pm0.88$ & 0.950\\
        & K $\alpha$ & 1.937 & $1.937\pm0.001$ & $10.40\pm1.08$ & 1.000\\
        & II - XI    &       &                 &                &      \\
     Ca & II-VII     & 3.359-3.352 & $3.356\pm0.005$ & $0.77\pm0.04$ & 1.015\\
     Ar & VI-IX      & 4.186-4.178 & $4.184\pm0.010$ & $0.82\pm0.04$ & 0.460\\
     S  & IX         & 5.320       & $5.322\pm0.018$ & $0.43\pm0.04$ & 0.046\\
        & IV-VIII    & 5.370-5.356 & $5.365\pm0.018$ & $1.26\pm0.06$ & 0.147\\
     Si & IX         & 6.947       & $6.937\pm0.004$ & $0.33\pm0.09$  & 0.017\\
        & VIII       & 7.007       & $7.008\pm0.005$ & $0.56\pm0.14$  & 0.036\\
        & VII        & 7.063       & $7.059\pm0.004$ & $0.75\pm0.14$  & 0.054\\
        & VI         & 7.117       & $7.117\pm0.005$ & $0.52\pm0.14$  & 0.042\\
        & II-VI      & 7.126-7.122 & $7.124\pm0.005$ & $0.74\pm0.13$  & 0.048\\
     Mg & V          & 9.814       & $9.827\pm0.011$ & $0.15\pm0.07$  & 0.0097\\
        & II-IV      & 9.890-9.883 & $9.893\pm0.008$ & $0.17\pm0.08$  & 0.0115\\
\hline
\vspace*{0.02in}
\end{tabular}
\parbox{3.2in}{
% \parbox{2.5in}{
\small\baselineskip 9pt
\footnotesize
\indent
$\rm ^1${from House 1969} \\
$\rm ^2${in units of $10^{-5}$ photons cm$^{-2}$s$^{-1}$} \\
$\rm ^3${fraction of column density with respect to Fe K (uncertainties are $\sim 15\%$, in case of Mg $45\%$} \\
}
\end{center}
\setcounter{table}{1}
\normalsize
\centerline{}
}

\subsection{Radiative Recombination Continua (RRC)}

In photoionized plasmas, free electrons are captured by ions, which result in
a continuous emission feature above the recombination edge (for a review, see
Liedahl et al. 2001 and references therein).  Since the plasma is overionized
relative to the local electron temperature, the widths of these continua are
narrow and appear as line-like features in the spectrum.  RRCs from
\ion{Ne}{10} and \ion{Ne}{9} at 9.10 \AA\ and 10.37 \AA, respectively, are
clearly observed.  Although some of the features are rather weak, they are consistent
with kT =  $\sim 10\pm2 ~\rm{eV}$ (for Ne X) or a temperature of $T \sim 1.2 \times 10^5
~\rm{K}$.  The \ion{Mg}{11} RRC is blended with the Si fluorescence lines, and
a possible RRC of \ion{O}{8} is observed at $\sim 14.25$ \AA.  The detection
of such narrow RRCs clearly indicates that there is a warm photoionized medium
associated with the wind.

For \ion{Ne}{10} at 12.13~\AA~ and 9.10~\AA~, the measured Ly$\alpha$/RRC flux ratio is $\sim 2.7 \pm
0.58$ after correcting for interstellar column density, and may be higher if
there is additional absorption above the Galactic value. Liedahl and Paerels (1996) predict for the
optically thin limit a ratio of 1.3 for a purely recombining plasma. Since this is inconsistent
with the observed ratio, we assume that the assumption of a purely recombining plasma is not correct here.
A more recent calculation by Kinkhabwala et al. (2001) also includes photoexitations in a more
self-consistent manner and predicts
a  ratio of Ly$\alpha$/RRC $= 2.7$ at $kT = 10 ~\rm{eV}$ 
which agrees well with our observations. The spectrum also exhibits
very weak, but detectable amounts of line flux from Fe L ions.  This is consistent
with the prediction by Kallman et al.\ (1996) that recombination emission from
K-shell ions dominates the spectrum in an overionized plasma.

\subsection{H- and He-like Lines}

The spectrum shown in Figure 2 also shows very strong emission lines (blue
model) from H- and He-like S, Si, Mg, Ne, and possibly O.  The measured line
fluxes range between 0.2 and $3 \times 10^{-5} ~\rm{photons~cm}^{-2}
~\rm{s}^{-1}$, and appear unresolved with no significant systematic line
shifts.  In the case of
\ion{Ne}{10}, four lines from the Lyman series are detected, with
Ly$\alpha$:$\beta$:$\gamma$:$\delta$ line ratios of $1.0$:$0.32$:$0.24$:$0.10$
(with uncertainties of the order of 25$\%$).  The
fluxes of higher lines in the Lyman series are larger than those expected from either a
purely recombining plasma or a purely collisionally ionized plasma, and
implies the presence of photoexcitation. This is consistent with the high
Ly$\alpha$/RRC flux ratio found in the previous section. For a photoionized
plasma including the effects of photoexcitation self-consistently, the
theoretical ratios for \ion{Ne}{10} at $kT = 10 ~\rm{eV}$ are
$1.0:0.35:0.16:0.08$ at a column density of $N_{\rm Ne X} = 3 \times 10^{17}
~\rm{cm}^{-2}$ and a turbulent velocity of $300 ~\rm{km~s}^{-1}$ (Kinkhabwala
et al.\ 2001).  The Lyman series for \ion{Mg}{12} is difficult to assess
since the Ly$\beta$ and $\gamma$ appear blended.  

The He-like lines of \ion{Si}{13} and \ion{Mg}{11} are shown in Figure 2 (middle and right). 
In both cases the resonance and forbidden line fluxes have roughly equal  
strengths, which is not expected in a predominantly low density recombining plasma,
but consistent with the fact that effects from photoexcitation have to be considered.
It would therefore be incorrect to interprete these triplets using the calculations by Porquet and Dubau (2000),
which in fact would predict a hot and dense collisional-dominated plasma. Such an interpretation would be
in contrast with the results above, which indicate the presence of cool and warm 
photoionized media through the detection of narrow RRC and moderate line strength.
Here Wojdowski et al. (2001) suggest that resonance scattering is the most probable
probable process to enhance the resonance line in the He-like triplets.

\centerline{\epsfxsize=8.5cm\epsfbox{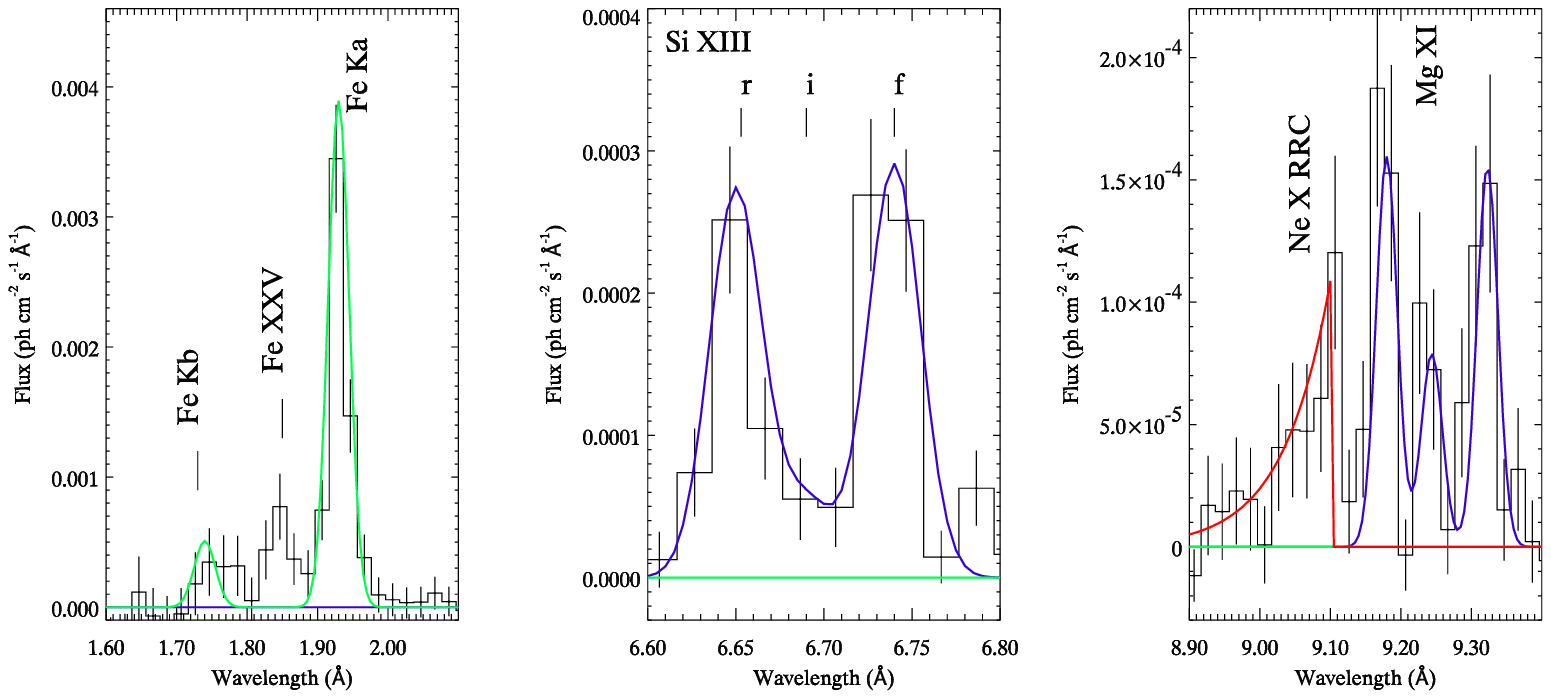}}
\figcaption{The Fe-K region (left) includes Fe K $\alpha$ and $\beta$ line emission as well as
a highly ionized Fe XXV line. The Si XIII He-like triplet (middle) shows a dominating resonance
and forbidden line. The NE X RRC (right) appears at 9.10 ~\AA~ below the Mg He-like triplet. Its
fit indicates a temperature of $\sim 10$ eV. 
\label{counts}}

We have yet to include contributions from the companion wind itself. 
In this case we have to consider collisionally ionized plasmas. Using
the Astrophysical Plasma Emission Code (APEC, Smith et al. 2001), we can calculate
line ratios between the Si XIII resonance line and Fe L lines at representative
temperatures. Temperatures in O-stars range between $\sim 5\times 10^6$ K (as measured in $\zeta Ori$,
(Waldron and Cassinelli 2001) and in an extreme case 5$\times 10^7$ K (as measured in   
$\theta$ Ori C, Schulz et al. 2000) from highly resolved X-ray lines. For now we assume that
this is also a representative range of possible temperatures in the winds of B0.5 Ib supergiants.
In the high temperature case we can exclude a significant contribution,
because here Fe XXIV lines should dominate the spectrum, which we do not observe. Similarly,
in the low temperature case Fe XVII lines should dominate the spectrum, which we do not
observe either. For the intermediate range we estimate, under the assumption 
that the sporadic and weak detections of Fe L lines
indeed reflect the existence of collisionally ionized emission, 
a contribution of a few percent at most.
 
\section{Conclusions}

The X-ray spectrum of Vela X-1 during eclipse shows a large variety of emission features
including lines from H-like and He-like ions, RRCs, a population of very weak Fe L lines,
and a number of fluorescence emission lines from mid-$Z$ elements and iron indicating a complex
structure of the X-ray emitting plasma. 

From the fit of the shape of the Ne X RRC at 9.10~\AA~ we deduce a temperature of $1.2 \times 10^5
~\rm{K}$ for the plasma. Its flux ratio with the Ne X Lyman $\alpha$ line is
consistent with an optically thin photoionized plasma. The spectrum shows clear evidence
of photoexcitation processes, for which resonance scattering from the strong eclipsed 
X-ray continuum may be a likely source. More detailed calculations on this issue 
are required.      
 
Fluorescence lines from various charge states were detected and partially identified. They
include strong emission from M-shell ions, but also from various L-shell ions. 
Fluorescence yields are generally extremely low in mid-$Z$ elements even at moderate
optical depths. The correlation of the inferred column density with Z implies that autoionization
dominates the fluorescence yields in these ions and the fluorescent plasma is not optically thin.
If the fluorescent plasma coexists with an optically thin photoionized plasma or in a region outside
the warm photoionized region, its has to be in the form of dense cool clumps.
Both scenarios may be the case, which is not inconsistent with current stellar wind models. For example,
Feldmeier (1995) pointed out that the wind can be devided into two distinct regions, an active
inner one where shell to shell collisons occur and an outer, quiescent region with "old" material.

The detection of fluorescence from L-shell ions is still mysterious. Given the relative low
fluorescent yields for these ions, resonant Auger destruction would have almost completely destroyed
these lines. Further investigations of the possible mechanisms and physical conditions that prevent the
lines from being destroyed are required. 

\acknowledgments The authors want to thank the MIT HETG team and the Chandra
X-ray Center for their support, and Ali Kinkhabwala for providing results
from calculations of photoionized plasmas in progress. This
research is funded by contracts SV-61010 and NAS8-39073.

\end{document}